# Conjectured Metastable Super-Explosives formed under High Pressure for Thermonuclear Ignition

**F. Winterberg**

University of Nevada Reno

Reno, Nevada

## Abstract

If matter is suddenly put under a high pressure, for example a pressure of 100 Mb $=10^{14}$ dyn/cm$^2$, it can undergo a transformation into molecular excited states, bound by inner electron shells, with keV potential well for the electrons. If this happens, the electrons can under the emission of X-rays go into the groundstate of the molecule formed under the high pressure. At a pressure of the order $\sim 10^{14}$ dyn/cm$^2$, these molecules store in their excited states an energy with an energy density of the order $\sim 10^{14}$ erg/cm$^3$, about thousand times larger than for combustible chemicals under normal pressures. Furthermore, with the much larger optical path length of keV photons compared to the path length of eV photons, these superexplosives can reach at their surface an energy flux density ($c=3\times10^{10}$ cm/s) of the order $(c/3)\times10^{14} = 10^{24}$ erg/cm$^2$s $= 10^{17}$ W/cm$^2$, large enough for the ignition of thermonuclear reactions.

## 1. Introduction

Under normal pressure the distance of separation between two atoms in condensed matter is typically of the order $10^{-8}$ cm, with the distance between molecules formed by the chemical binding of atoms of the same order of magnitude. As illustrated in a schematic way in Fig. 1, the electrons of the outer electron shells of two atoms undergoing a chemical binding form a "bridge" between the reacting atoms. The formation of the bridge is accompanied in a lowering of the electric potential well for the outer shell electrons of the two reacting atoms, with the electrons feeling the attractive force of both atomic nuclei. Because of the lowering of the potential well, the electrons undergo under the emission of eV photons a transition into lower energy molecular orbits.

Going to still higher pressures, a situation can arise as shown in Fig.2, with the building of electron bridges between shells inside shells. There the explosive power would be even larger. Now consider the situation where the condensed state of many closely spaced atoms is put under high pressure making the distance of separation between the atoms much smaller, whereby the electrons from the outer shells coalesce into one shell surrounding both nuclei, with electrons from inner shells forming a bridge. Because there the change in the potential energy is much larger, the change in the electron energy levels is also much larger, and can be of the order of keV. There then a very powerful explosive is formed releasing its energy in a burst of keV X-rays. This powerful explosive is likely to be very unstable, but it can be produced by the sudden application of a high pressure in just the moment when it is needed. Because an intense burst of X-rays is needed for the ignition of a thermonuclear microexplosion, the conjectured effect if it exists, has the potential to reduce the cost for the ignition of thermonuclear microexplosions by orders of magnitude.

## 2. An instructive example

The energy of an electron in the groundstate of a nucleus with the charge Ze is

$$E_1 = -13.6 Z^2 \text{ [eV]}. \qquad (1)$$

With the inclusion of all the Z electrons surrounding the nucleus of charge Ze, the energy is

$$E_1^* \approx -13.62 Z^{2.42} \text{ [eV]} \qquad (2)$$

with the outer electrons less strongly bound to the nucleus.

Now, assume that two nuclei are so strongly pushed together that they act like one nucleus with the charge 2Ze, onto the 2Z electrons surrounding the 2Ze charge. In this case, the energy for the innermost electron is

$$E_2 = -13.6(2Z)^2 \text{ [eV]} \tag{3}$$

or if the outer electrons are taken into account,

$$E_2^* = -13.6(2Z)^{2.42} \text{ [eV]} \tag{4}$$

For the difference one obtains

$$\delta E = E_1^* - E_2^* = 13.6Z^{2.42}(2^{2.42} - 1) \approx 58.5Z^{2.42} \text{ [eV]}. \tag{5}$$

Using the example Z=10, which is a neon nucleus, one obtains $\delta E \approx 15$ keV. Of course, it would require a very high pressure to push two neon atoms that close to each other, but this example makes it plausible that smaller pressures exerted on heavier nuclei with many more electrons may result in a substantial lowering of the potential well for their electrons.

## 3. Several ways to reach a pressure of 100 Mb

A pressure of $p \approx 100$Mb = $10^{14}$ dyn/cm$^2$, can be reached with existing technology in sufficiently large volumes, with at least three possibilities:

1. Bombardment of a solid target with an intense relativistic electron- or ion beam.
2. Hypervelocity impact.
3. Bombardment of a solid target with beams or by hypervelocity impact, followed by a convergent shock wave.

To 1: This possibility was considered by Kidder [1] who computes a pressure of 50 Mb, if an iron plate is bombarded with a 1 MJ – 10 MeV – $10^6$ A relativistic electron beam, focused down to an area of 0.1 cm$^2$. Accordingly, a 2 MJ beam would produce 100

Mb. Instead of using an intense relativistic electron beam, one may use an intense ion beam. It can be produced by the same high voltage technique, replacing the electron beam diode by a magnetically insulated diode [2].

Using intense ion beams has the additional benefit that the stopping of the ions in a target is determined by a Bragg curve, generating the maximum pressure inside the target, not on its surface.

To 2: A projectile with the density $\rho \approx 20\,\mathrm{g/cm^3}$, accelerated to a velocity v = 30 km/s would, upon impact, produce a pressure of $p \approx 100\,\mathrm{Mb}$. The acceleration of the projectile to these velocities can be done by a magnetic traveling wave accelerator [3, 4].

To 3: If, upon impact of either a particle beam or projectile, the pressure is less than 100 Mb, for example only of the order 10 Mb, but acting over a larger area, a tenfold increase in the pressure over a smaller area is possible by launching a convergent shock wave from the larger area on the surface of the target, onto a smaller area inside. According to Guderley [5], the rise in pressure in a convergent spherical shock wave goes as $r^{-0.9}$, which means that 100 Mb could be reached by a ten-fold reduction in the radius of the convergent shock wave.

While it is difficult to reach 30 km/s with a traveling magnetic wave accelerator, it is easy to reach a velocity of 10km/s with a two stage light gas gun.

## 4. Equation of state for matter under a pressure of 100Mb

We assume an equation of state of the form $p/p_o = (n/n_o)^\gamma$. For a pressure of 100Mb = $10^{14}$ dyn/cm$^2$, we may set $\gamma = 3$ and $p_o = 10^{11}$ dyn/cm$^2$, $p_o$ being the Fermi pressure of a solid at the atomic number density $n_o$, with $n$ being the atomic number density at the elevated pressure $p > p_o$. With $d = n^{-1/3}$, where $d$ is the lattice constant, one has

$$d/d_o = (p_o/p)^{1/9} \tag{6}$$

For $p = 10^{14}$ dyn/cm$^2$. Such a lowering of the inneratomic distance is sufficient for the formation of molecular states.

## 5. Emission of X-rays under the sudden application of a high pressure pulse

Calculations done by Muller, Rafelski, and Greiner [6] (see Fig. (4-6) in Appendix A), show that for molecular states ${}_{35}$Br- ${}_{35}$Br, ${}_{53}$I- ${}_{79}$Au, and ${}_{92}$U-${}_{92}$U, a two-fold lowering of the distance of separation leads to a lowering of the electron orbit energy eigen values by ~ 0.35 keV, 1.4 keV, respectively. At a pressure if 100 Mb = $10^{14}$ dyn/cm$^2$ where $d/d_0 \cong 1/2$, the result of these calculations can be summarized by ($\delta$E in keV)

$$\log \delta E \cong 1.3\times 10^{-2} Z - 1.4 \tag{7}$$

replacing eq (5), where Z is here the sum of the nuclear charge for both components of the molecule formed under the high pressure.

The effect the pressure has on the change in these quasi-molecular configurations is illustrated in Fig. 3, showing a $p - d$ (pressure-lattice distance) diagram. This diagram illustrates how the molecular state is reached during the compression along the adiabat $a$ at the distance $d = d_c$ where the pressure attains the critical value $p = p_c$. In passing over this pressure the electrons fall into the potential well of the two-center molecule, releasing their potential energy as a burst of X-rays. Following its decompression, the molecule disintegrates along the lower adiabat $b$.

The natural life time of an excited atomic (or molecular) state, emitting radiation of the frequency ν is given by [7]

$$\tau_s \cong 3.95 \times 10^{22} / \nu^2 \text{ [sec]} \tag{8}$$

For keV photons one finds that $\nu \cong 2.4 \times 10^{17}$ s$^{-1}$, and thus $\tau_s \cong 6.8 \times 10^{-14}$ sec.

By comparison, the shortest time for the high pressure rising at the front of a shock wave propagating with the velocity v through a solid with the lattice constant d, is of the order

$$\tau_c \cong d / \mathrm{v} \tag{9}$$

Assuming that $\mathrm{v} \cong 10^6$ cm/s, a typical value for the shock velocity in condensed matter under high pressure, and that $d \cong 10^{-8}$ cm, one finds that $\tau_c \cong 10^{-14}$ sec. In reality the life time for an excited state is much shorter than $\tau_s$, and of the order of the collision time, which here is the order of $\tau_c$.

The time for the electrons to form their excited state in the molecular shell, is of the order $1/\omega_p \sim 10^{-16}$ sec, where $\omega_p$ is the solid state plasma frequency. The release of the X-rays in the shock front is likely to accelerate the shock velocity, exceeding the velocity profile of the Guderley solution for the convergent shock waves.

A problem for the use of these contemplated super-explosives to ignite thermonuclear reactions is the absorption of the x-ray in dense matter. It is determined by the opacity [8]

$$\kappa = 7.23 \times 10^{24} \rho T^{-3.5} \sum_i \frac{w_i Z_i^2}{A_i} \frac{g}{t} \tag{10}$$

Where $w_i$ are the relative fractions of the elements of charge $Z_i$ and atomic number $A_i$ in the radiating plasma, with g the Gaunt and t the guilliotine factor.

The path length of the x-ray is then given by

$$\lambda = (\kappa \rho)^{-1} \tag{11}$$

This clearly means that in material with a large $Z$ value, the path length is much smaller than for hydrogen where $Z$ = 1, This suggests to place the super-explosive in a matrix of particles, thin wires or sheets embedded in solid hydrogen. If the thickness of the particles, thin wires or sheets is smaller than the path length in it for the X-ray , the X-ray can heat up to high temperature the hydrogen, if the thickness of the surrounding hydrogen is large enough for the X-ray to be absorbed in the hydrogen. The hydrogen is thereby transformed into a high temperature plasma, which can increase the strength of the shock wave generating the X-ray releasing pressure pulse.

## 6. The X-ray flux released by the high pressure

If the change in pressure is large, whereby the pressure in the upper adiabat is large compared to the pressure in the lower adiabat, the X-ray energy flux is given by the photon diffusion equation

$$\phi = -\frac{\lambda c}{3} \nabla w \tag{12}$$

where $w$ is the work done per unit volume to compress the material, where $w = p/(\gamma - 1)$.

For $\gamma = 3$, one has $w = p/2$, whereby (12) becomes

$$\phi = -\frac{\lambda c}{6} \nabla p \tag{12a}$$

Assuming that the pressure e-folds over the same length as the photon mean free path, one has

$$\phi \sim (c/6)\,p \tag{13}$$

For the example p = 100 Mb = $10^{14}$ dyn/cm$^2$ one finds that $\phi \sim 5\times10^{23}$ erg/cm$^2$s = $5\times10^{16}$ W/cm$^2$.

**7. How to find and prepare the best super-explosive**

If the conjectured super-explosive consists of just one element, as is the case for the $_{35}Br$ - $_{35}Br$ reaction, or the $_{92}U$ - $_{92}U$ reaction, no special preparation for the super-explosive is needed. But as the example of Al – FeO thermite reaction shows, reactions with different atoms can release a much larger amount of energy compared to other chemical reactions. For the conjectured super-explosives this means as stated above that they have to be prepared as homogeneous mixtures of nano-particle powders, bringing the reacting atoms come as close together as possible.

**8. Possible connection to the release of intense X-ray bursts in electric pulse power driven exploding wire arrays**

The outstanding unresolved question in the release of intense X-ray bursts in electric pulse power-driven imploding wire arrays is the unexpected occurrence of a large contribution in the keV energy range. It can not be explained by a simple kinetic into thermal energy conversion model, which predicts a black body radiation spectrum in the sub-keV range [9, 10].

To explore the possibility whether or not these X-ray bursts can be explained by the proposed mechanism, we take as an example a thin wire with a radius $r = 2\times10^{-3}$ cm over which a current of I = $10^6$ A is flowing. The magnetic field at the surface of the wire is there $B = 0.2\,I/r = 10^8\,G$, with a magnetic pressure $B^2/8\pi = 4\times10^{14}$ dyn/cm$^2$, acting on the wire, sufficiently large for the conjectured release of keV X-rays by 100 Mb pressures. Unexplained X ray bursts have also been observed by the high velocity impact of large molecules or clusters, accelerated to a velocity in excess of 100 km/s. At these impact velocities pressures in the 100 Mb range can be expected [11].

## 9. Thermonuclear Ignition

For the ignition of a thermonuclear reaction one may consider the following scenario illustrated in Fig. 3. A convergent shock wave launched at the radius $R = R_0$ into a spherical shell of outer and inner radius $R_0$, $R_1$, reaches near the radius $R = R_1$ at a pressure of 100 Mb. After the inward moving convergent shock wave has reached the radius $R = R_1$, an outward moving rarefaction wave is launched from the same radius $R = R_1$, from which an intense burst of X-rays is emitted. One can then place a thermonuclear target inside the cavity of the radius $R = R_1$, with the target bombarded, imploded, and ignited by the X-ray pulse.

Instead of aiming at the release of the X-rays in the pressure spike of a shock wave, one may isentropically compress the super-explosive by a programmed pressure pulse, until the moment it reaches the critical pressure $p_c$, where the X-rays are explosively released. However, with the super-explosive made up from high *Z*-value atoms, the X-rays will be entrapped inside the super-explosive, with the result that the X-ray radiation energy is converted into black body radiation. To prevent this from happening, one may place a matrix of small particles made up from the super-explosive in solid hydrogen. If the particles are small enough to be transparent with regard to the X-rays released, the X-rays will heat up the surrounding hydrogen to high temperatures. There then, the energy released by the super-explosive will be transformed into thermal kinetic energy of the high temperature hydrogen, which in turn can be used for the implosion of a thermonuclear target.

**Conclusion**

The ignition of thermonuclear micro-explosions requires an energy of more than 1 MJ ($10^{13}$ erg), to be delivered onto an area less than one $cm^2$ in less than $10^{-8}$ sec. This energy is really not that large, but difficult to deliver it in the short time by laser- or particle beams. Even at the required beam intensity, the particle number density in the beam is much smaller than the particle number density of the solid state. The only kind of beam where the number density is comparable is in impact fusion, where the beam is a single solid particle (marco-particle), accelerated to high velocities. There the energy flux density is $\phi = (1/2)\rho v^3$, where $\rho$ is the density of the macroparticle and v its velocity. For the example $\rho = 20\,\text{g/cm}^3$, a velocity of v = 50 km/s is needed to reach $\phi = 100\,\text{TW}$ ($10^{21}$ erg/s). This velocity is less than v = 200 km/s needed for impact fusion, but sufficient if the impact energy can be converted into a burst of radiation to implode and ignite a thermonuclear target. The conversion of the kinetic impact energy into heat, with the energy emitted as thermal radiation, is not very efficient if a large fraction of the heat is trapped inside the solid. This is different if the energy stored in the pressure is directly converted into X-rays, having a large range.

If realizable, the concept presented here opens new avenues for the achievement of nuclear fusion by inertial confinement. To reach a pressure of 100 Mb, one can use intense relativistic electron beams, which are less expensive by orders of magnitude than lasers. And in comparison to soft X-rays emitted by exploding wire arrays, the concept does not need replaceable transmission lines, because intense relativistic electron beams can with ease be transported through a tenuous background gas. The concept also sheds new light on fusion through the acceleration of macro-particles, because if coupled with the generation of X-rays by the impact pressure of the macro-particles, smaller velocities suffice than the velocities otherwise required for impact fusion.

**Acknowledgement**

The idea presented in this paper, first considered in 1974, was stimulated by the paper of B. Müller, J. Rafelski, and W. Greiner, and I acknowledge the very useful conversation I had at that time with Dr. Rafelski in Frankfurt, Germany.

**Appendix : The quantum mechanical eigenvalues of two-atom molecule two-center configurations**

This is the extension of the Heitler-London theory for the chemical bonding of two hydrogen atoms to heavier elements. It is fortunate that this problem has already been solved to predict the outcome of heavy atom collisions, and one can simply use the results obtained [6]. Because in the collision of the heaviest atoms, the effective two center electric charge can exceed the critical value $Z_{crit}$ = 137, above which vacuum breakdown begins to set in. One has here to use the two center Dirac equation for an electron of mass M in the field of two electric charges located at **–R** and **R**:

$$[c\boldsymbol{\alpha}\bullet\mathbf{p} + \beta Mc^2 - E + V_1(\mathbf{r} - \mathbf{R}) + V_2(\mathbf{r} + \mathbf{R})]\,\Psi = 0. \tag{A.1}$$

Introducing prolate spheroidal coordinates $\xi,\ \eta,\ \varphi$, one has with the z-axis going from –**R** to **R**:

$$\left.\begin{aligned} x &= R\left[\left(\xi^2-1\right)\left(1-\eta^2\right)\right]^{1/2}\cos\varphi \\ y &= R\left[\left(\xi^2-1\right)\left(1-\eta^2\right)\right]^{1/2}\sin\varphi \\ z &= R\xi\eta \end{aligned}\right\} \tag{A.2}$$

With $m+1/2$ the angular momentum around the z-axis, the $\varphi$ dependence in (A.1) can be separated setting

$$\Psi(\mathbf{r}) = \begin{pmatrix} e^{im\varphi} & & & \\ & e^{i(m+1)\varphi} & & \\ & & ie^{im\varphi} & \\ & & & ie^{i(m+1)\varphi} \end{pmatrix} \psi'(\xi,\eta) \tag{A.3}$$

resulting in

$$\left[\frac{\hbar c}{R\left(\xi^2-\eta^2\right)}\begin{pmatrix} 0 & 0 & \pi_z & \pi^- \\ 0 & 0 & \pi^+ & -\pi_z \\ -\pi_z & -\pi^- & 0 & 0 \\ -\pi^+ & \pi_z & 0 & 0 \end{pmatrix} + \beta \mathrm{M}c^2 - \mathrm{E} + V_1\left(\xi+\eta\right) + V_2\left(\xi-\eta\right)\right]\psi'(\xi,\eta) = 0 \qquad \text{(A.4)}$$

where

$$\left.\begin{aligned} \pi^+ &= \mathrm{W}\left(\xi\frac{\partial}{\partial\xi}-\eta\frac{\partial}{\partial\eta}\right)-\frac{m}{\mathrm{W}}\left(\xi^2-\eta^2\right) \\ \pi^- &= \mathrm{W}\left(\xi\frac{\partial}{\partial\xi}-\eta\frac{\partial}{\partial\eta}\right)+\frac{m+1}{\mathrm{W}}\left(\xi^2-\eta^2\right) \\ \pi_z &= \eta\left(\xi^2-1\right)\frac{\partial}{\partial\xi}+\xi\left(1-\eta^2\right)\frac{\partial}{\partial\eta} \\ \mathrm{W} &= \left[\left(\xi^2-1\right)\left(1-\eta^2\right)\right]^{1/2} \end{aligned}\right\} \qquad \text{(A.5)}$$

Eq. (A.4) is separable in a $\xi$ and $\eta$ dependence setting

$$\Psi_{nls}^{m}\left(\xi,\eta\right)=\left(\xi^2-1\right)^{(m+\varepsilon_s)/2} \times \exp\left(-\frac{\xi-1}{a}\right)\mathrm{L}_n^{m+\varepsilon_s}\left(\frac{\xi-1}{a}\right)\mathrm{P}_l^{m+\varepsilon_s}\left(\eta\right)\chi_s \qquad \text{(A.6)}$$

where $\mathrm{L}_\mathrm{i}^\alpha$ and $\mathrm{P}_\mathrm{i}^\alpha$ are the associated Laguerre and Legendre polynomials. In eq. (A.6), $a$ is a scaling factor, not affecting the energy eigenvalue E. Furthermore, $\chi_s$ are the unit spin vectors, $\varepsilon_s = 0$ for s odd and $\varepsilon_s = 1$ for s even. Inserting (A.6) into (A.4), one obtains the energy eigenvalues. As noticed by Muller, Rafelski, and Greiner [6], the energy eigenvalues exhibit a "run way" towards distances of separation of the order $10^3$ fm, which is an indication of the potential well in Fig. 3.

If the two atoms just touch each other without the application of an external pressure, one can approximately set for their distance of separation 1 Angström = $10^5$ fm.

Under the high pressure of $10^{14}$ dyn/cm$^2$, their mutual distance of separation is reduced from $10^5$ fm down to $5\times10^4$ fm.

**Figures:**

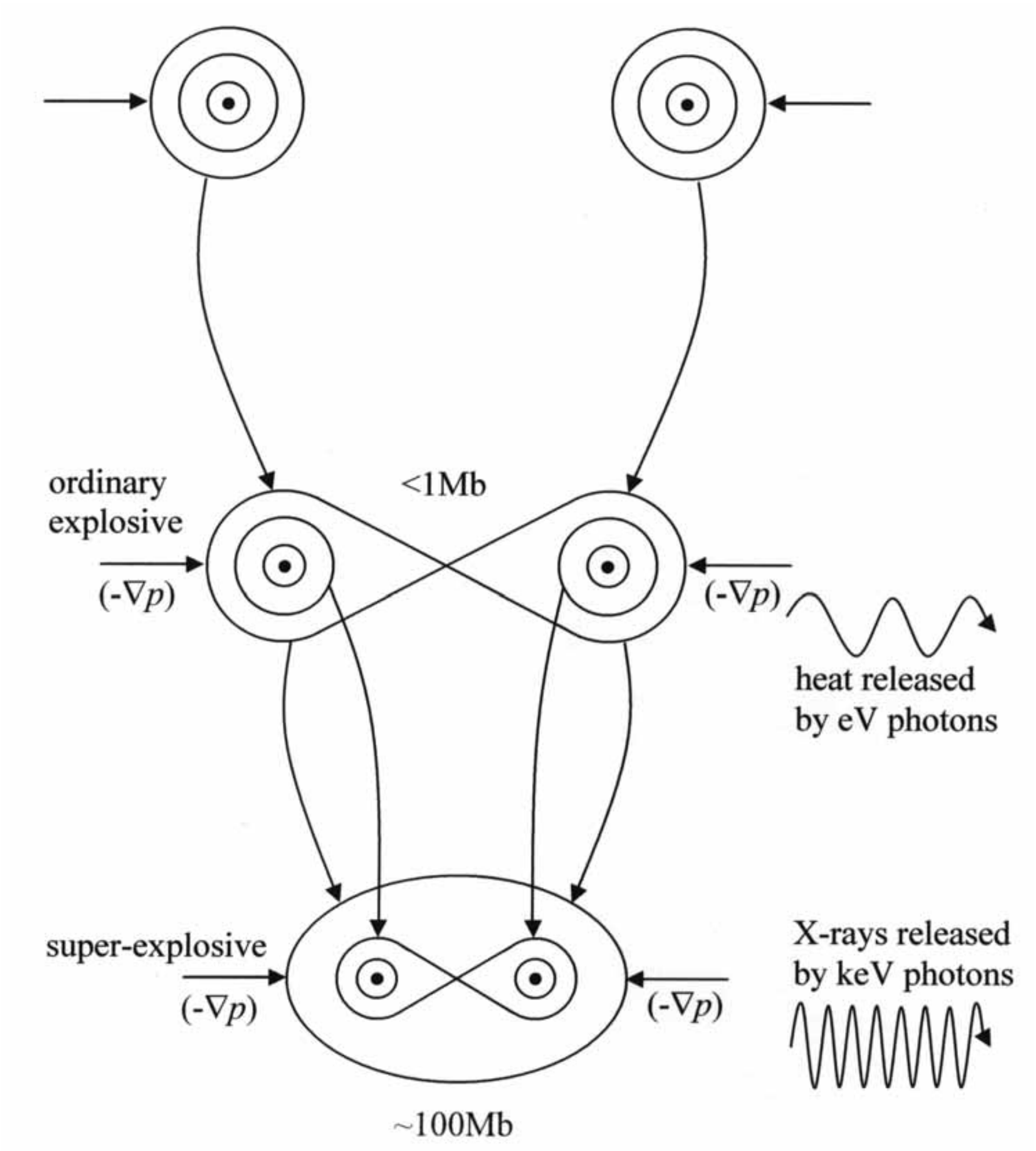

Fig.1. In an ordinary explosive the outer shell electrons of the reacting atoms form “eV” molecules accompanied by the release of heat through eV photons. In a superexplosive the outer shell electrons “melt” into a common outer shell with inner electron shells form “keV” molecules accompanied by the release of X-ray keV photons.

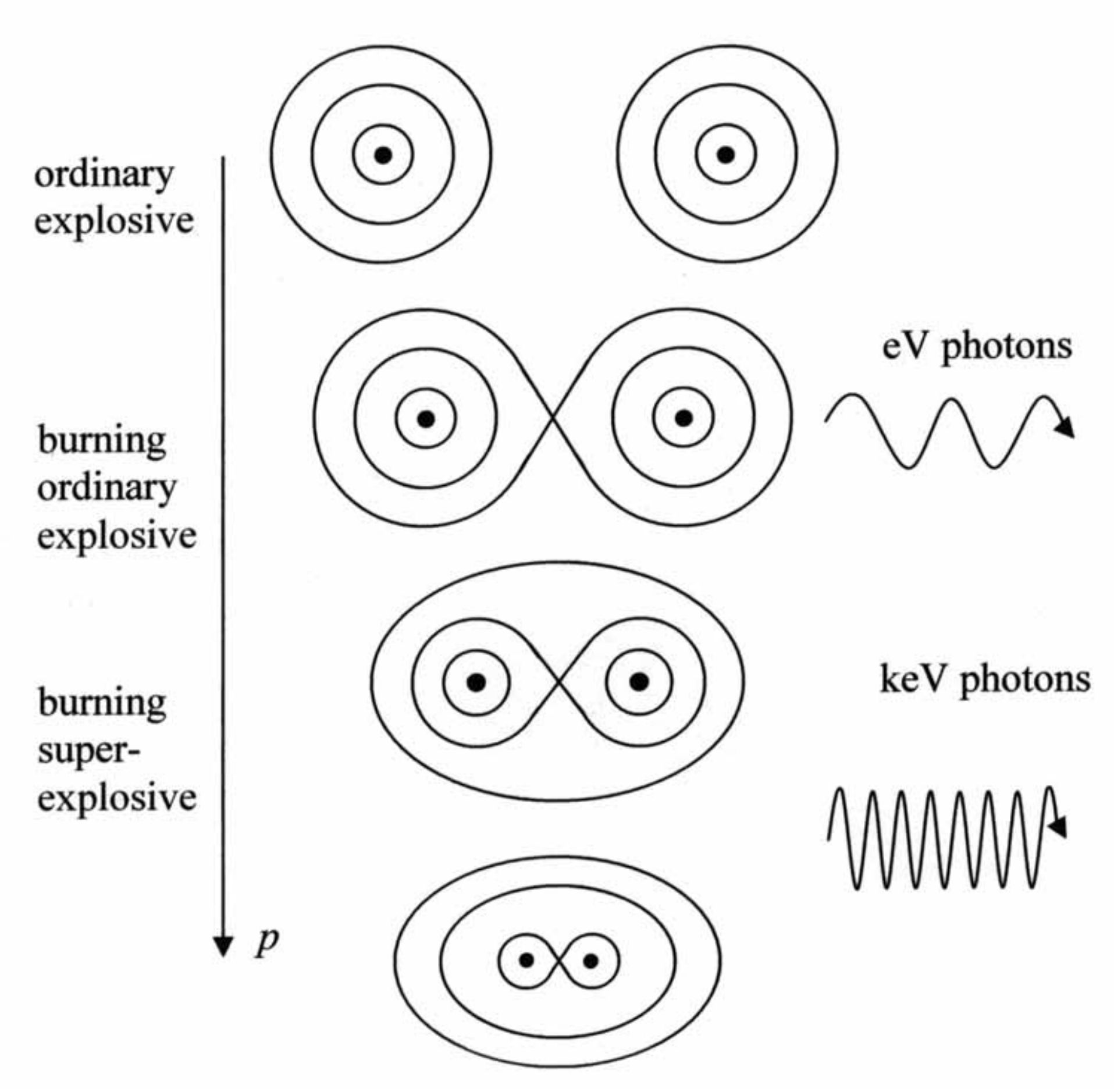

Fig. 2. With increasing pressure electron-bridges are formed between shells inside shells melting into common shells.

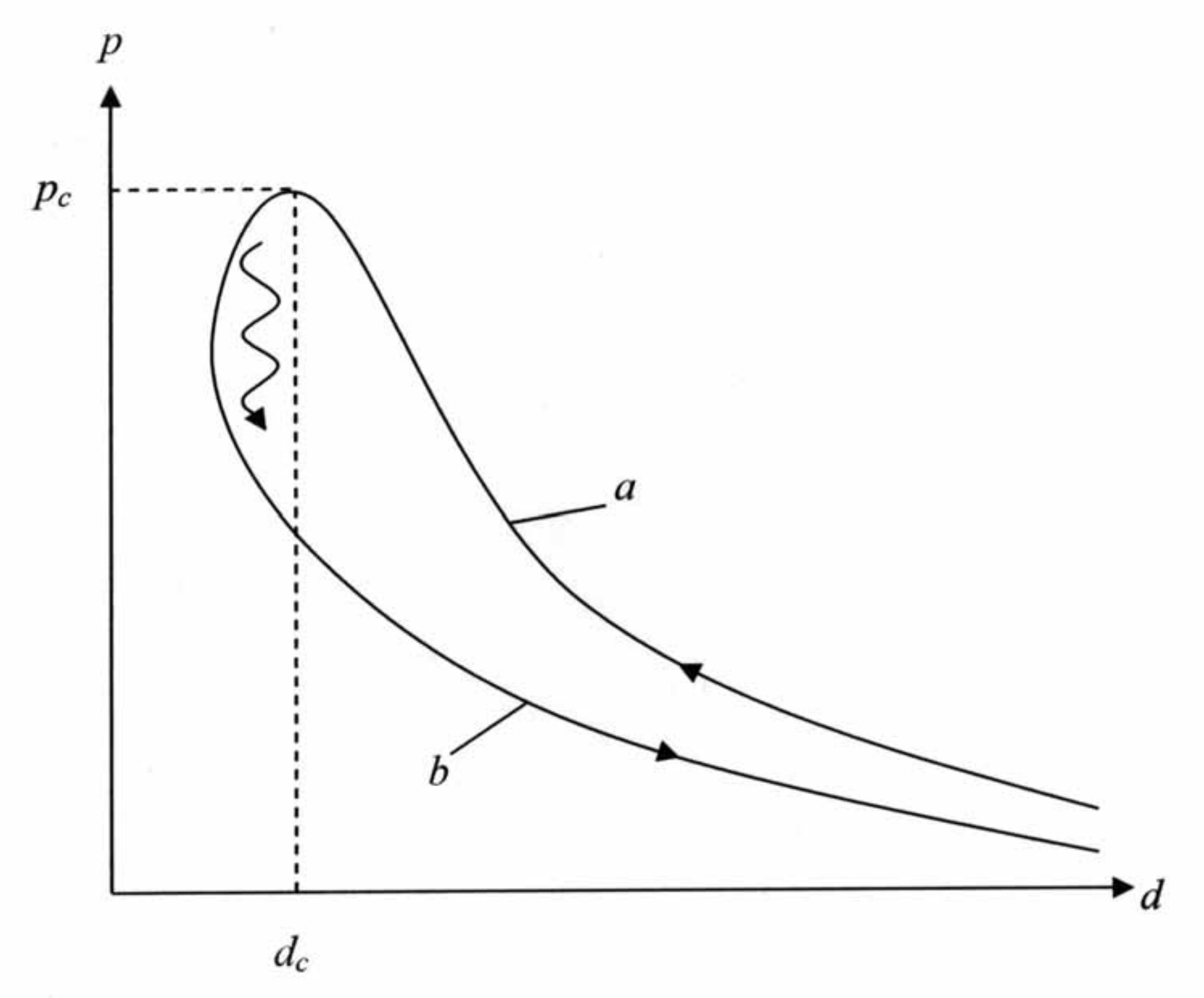

Fig. 3. *p-d,* pressure – inneratomic distance diagram for the upper atomic and lower molecular adiabat.

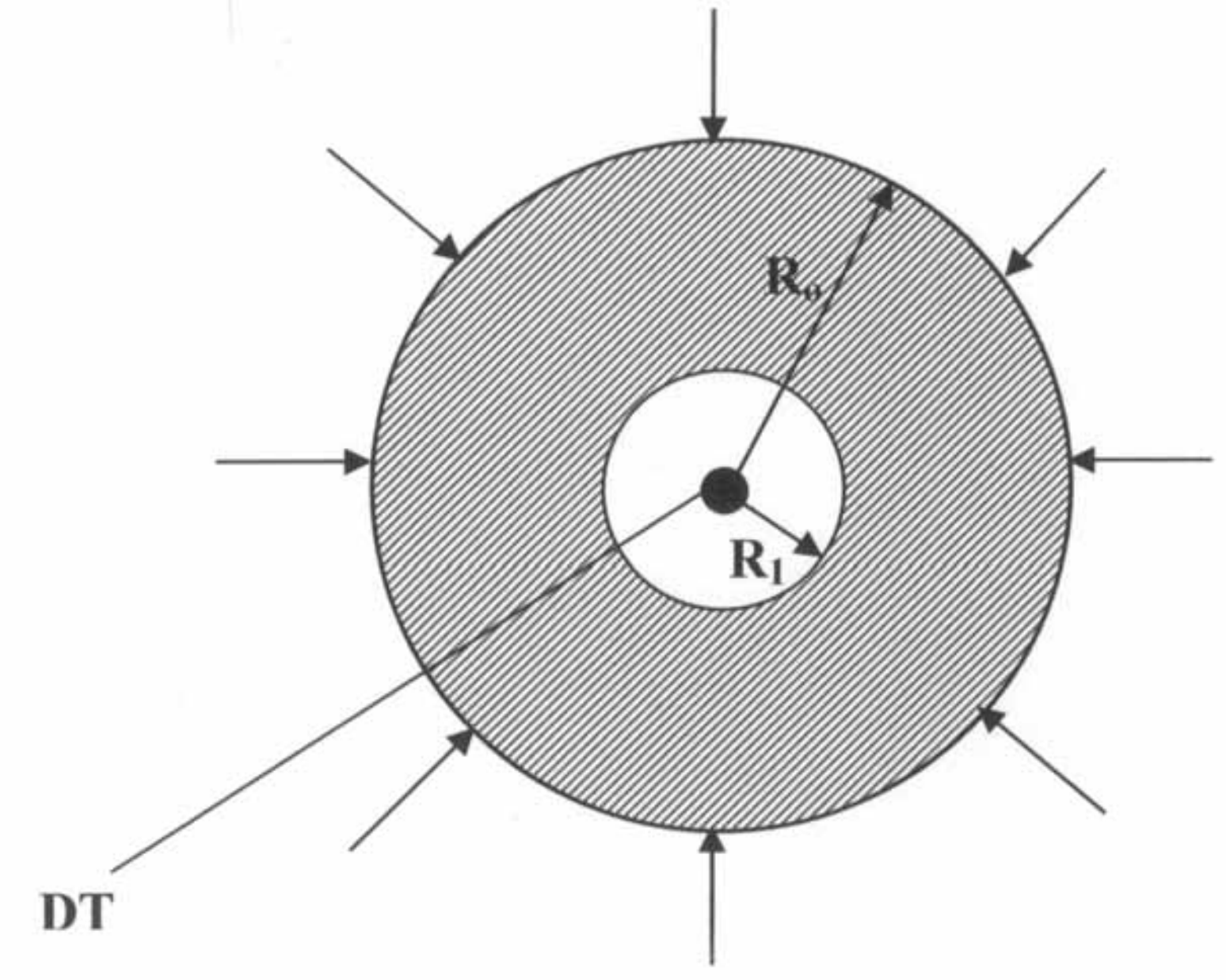

Fig. 4. Inertial confinement fast ignition configuration.